\documentclass[12pt]{article}
\usepackage{fancyhdr}

\usepackage{overcite}

\usepackage{graphicx}

\usepackage{amsfonts}

\usepackage{siunitx}

\usepackage{epstopdf}

\usepackage{caption}

\usepackage{subcaption}

\usepackage{authblk}

\usepackage[mathlines]{lineno}
\makeatletter \renewcommand\@biblabel[1]{$^{#1}$} \makeatother
\newlength{\bibhang}
\title{Heterogeneous multiscale Monte Carlo simulations for gold nanoparticle radiosensitization}
\author{Martin P. Martinov}
\author{Rowan M. Thomson}
\affil{Carleton Laboratory for Radiotherapy Physics, Department of Physics, Carleton University, Ottawa, Ontario, K1S 5B6, Canada}

\newcommand{\mfig}[1]{\marginpar{{\sf Fig~\ref{#1} }}}
\newcommand{\mtab}[1]{\marginpar{{\sf Table~\ref{#1} }}}
\newcommand{\ie}{{\it i.e.}, }
\newcommand{\eg}{{\it e.g.}, }

\newcommand{\cen}[1]{\begin{center} #1 \end{center}}

\newcommand{\captionl}[1]{\parbox{16.5cm}{\caption[dummy]{{\sf #1}}}}

\newcommand{\model}{heterogeneous multiscale } 

\newcommand{\acro}{HetMS }
\newcommand{\acroc}{HetMS}

\begin{document}

\cen{\sf {\Large {\bfseries Heterogeneous multiscale Monte Carlo simulations for gold nanoparticle radiosensitization} \\
Martin P. Martinov$^\text{a)}$ and Rowan M. Thomson$^\text{b)}$} \\ \vspace{3mm}
email: a) martinov@physics.carleton.ca and b) rthomson@physics.carleton.ca \\ \vspace{5mm}
Carleton Laboratory for Radiotherapy Physics, Department of Physics, Carleton University, Ottawa, Ontario, K1S 5B6, Canada \\ \vspace{5mm} 
Published in Medical Physics Volume 44, Issue 2 in February 2017 (DOI:10.1002/mp.12061) \\
}
\pagestyle{empty}

%           _         _                  _   
%     /\   | |       | |                | |  
%    /  \  | |__  ___| |_ _ __ __ _  ___| |_ 
%   / /\ \ | '_ \/ __| __| '__/ _` |/ __| __|
%  / ____ \| |_) \__ \ |_| | | (_| | (__| |_ 
% /_/    \_\_.__/|___/\__|_|  \__,_|\___|\__|
\begin{abstract}

\noindent\textbf{Purpose:}
To introduce the \model (\acroc) model for Monte Carlo simulations of gold nanoparticle dose-enhanced radiation therapy (GNPT), a model characterized by its varying levels of detail on different length scales within a single phantom; to apply the \acro model in two different scenarios relevant for GNPT and to compare computed results with others published.\\
%compare results with others published.   which is characterized by its varying levels of detail on different length scales within a single phantom.  \\%and apply the \model (\acroc) model within the context of gold nanoparticle dose-enhanced radiation therapy (GNPT).  The \acro model is a Monte Carlo simulation framework characterized by its varying levels of detail on different length scales within a single phantom.  
%The model is applied to two different scenarios relevant to GNPT, and results are compared to others published.\\
%OLD: This work introduces the \model (\acroc) model and applies it to gold nanoparticle dose-enhanced radiation therapy (GNPT).  The \acro model is a Monte Carlo simulation framework characterized by its varying levels of detail on different length scales within a single phantom.  The model is applied to two different scenarios relevant to GNPT, and compared to other published results.\\
\noindent\textbf{Methods:}
%\acro MC simulations are used to determine dose enhancement factors (DEFs) in multiple microcavities (in which GNPs are modeled discretely) situated within a larger scatter phantom (comprised of a homogeneous blend of water/tissue and gold).
The \acro model is implemented using an extended version of the EGSnrc user-code egs\_chamber; the extended code is tested and verified via comparisons with recently-published data from independent GNP simulations.  Two distinct scenarios for the \acro model are then considered: (1) monoenergetic photon beams (20~keV to 1~MeV) incident on a cylinder (1~cm radius, 3~cm length); %containing 5, 10, or 20~mg of GNPs per gram of tissue; %, considering concentrations of 5, 10, and 20 mg of Au per gram of tissue.
(2) isotropic point source (brachytherapy source spectra) at the center of a 2.5~cm radius sphere with gold nanoparticles (GNPs) diffusing outwards from the center. % to create a gradient in gold concentration decreasing with distance from the center.  
Dose enhancement factors (DEFs) are compared for different source energies, depths in phantom, gold concentrations, GNP sizes, and modeling assumptions, as well as with independently published values.  Simulation efficiencies are investigated.  \\
\noindent\textbf{Results:}
The \acro MC simulations account for the competing effects of photon fluence perturbation (due to gold in the scatter media) coupled with enhanced local energy deposition (due to modeling discrete GNPs within subvolumes).  DEFs are most sensitive to these effects for the lower source energies, varying with distance from the source; DEFs below unity (\ie dose decreases, not enhancements) can occur at energies relevant for brachytherapy.  For example, in the cylinder scenario, the 20 keV photon source has a DEF of 3.1 near the phantom's surface, decreasing to less than unity by 0.7 cm depth (for 20 mg/g).  Compared to discrete modeling of GNPs throughout the gold-containing (treatment) volume, efficiencies are enhanced by up to a factor of 122 with the \acro approach.
For the spherical phantom, DEFs vary with time for diffusion, radionuclide, and radius; DEFs differ considerably from those computed using a widely-applied analytic approach.  \\  %; discrepancies with published DEFs are discussed.
%In the cylinder scenario, the DEF in the scoring region varies by a factor of 3 depending on the depth of the scoring region for the 20 keV beam at a concentration of 5 mg/g.  At a concentration of 20 mg/g and a beam energy of 30 keV, the DEF is as high as 4 near the surface of the cylinder and drops to below 1 as depth increases.  The variance of DEF as a function of depth is smaller for larger energies, DEF is constant at energies at or above 200 keV.  In the sphere scenario, the dose within 7 mm of the source in the gold case increases or remains the same as the water case.  At distances above 1.2 cm, the DEF is at or below unity, depending on the time the gold has had to diffuse.\\
\noindent\textbf{Conclusions:}
By combining geometric models of varying complexity on different length scales within a single simulation, the \acro model can effectively account for both macroscopic and microscopic effects which must both be considered for accurate computation of energy deposition and DEFs for GNPT.  Efficiency gains with the \acro approach enable diverse calculations which would otherwise be prohibitively long. % and not possible  due to limited computing power.
The \acro model may be extended to diverse scenarios relevant for GNPT, providing further avenues for research and development.
%There are both microscopic and macroscopic effects that affect DEF in a microscopic region, which must be accounted for to properly estimate DEFs expected in GNPT.  With more careful modelling and validation, accounting for the variance demonstrated in this work, future GNPT models can begin to provide clearer picture of the effects of GNPs in radiation therapy. 

\end{abstract}

\newpage
% Good for reviewing
\setlength{\baselineskip}{0.7cm}
\pagestyle{fancy}

%  _____       _             
% |_   _|     | |            
%   | |  _ __ | |_ _ __ ___  
%   | | | '_ \| __| '__/ _ \ 
%  _| |_| | | | |_| | | (_) |
% |_____|_| |_|\__|_|  \___/                             
\section{Introduction}

The pioneering work of Hainfield \textit{et al} using gold nanoparticles (GNPs) during irradiation of tumor-bearing mice to increase survival\cite{Ha04c} has sparked a large number of studies investigating the potential of GNPs in human radiotherapy treatments.  While experimental work investigates various metrics related to the development of GNP dose-enhanced radiation therapy (GNPT), numerical methods are employed to investigate radiation transport and energy deposition due to the limitations of experimental dosimetry in this context.  Diverse computational models are employed to study increased energy deposition due to gold nanoparticles, from analytic techniques\cite{Ng10,Ng12b,Ng12a} to Monte Carlo (MC) simulations\cite{Jo10,Do13,Le11c}.  In particular, computations are often used to determine the relative increase in dose due to the presence of GNPs, defined as the dose enhancement factor (DEF), \ie the ratio of dose with GNPs present to that without.

Accurate computations of DEFs in GNPT can present considerable challenges.  For realistic radiation therapy scenarios, energy deposition within both the treatment volume and surrounding normal healthy tissues must be considered, resulting in volumes of interest greater than (1~cm)$^3$.  For a frequently-cited concentration of gold (7 mg of gold per g of tissue or water), there may be up to $\sim10^{16}$ GNPs in (1~cm)$^3$ of tissue, depending on GNP size.  Modeling such a large number of GNPs strains most computational approaches.  Within the context of MC simulations, most studies considering whole tumor volumes of interest (millimetres or greater) use a macroscopic model in which voxels represent a homogeneous mixture of soft tissue and gold with a single mass density\cite{Ch05l,Ch09d} to avoid modeling large numbers of GNPs.  However, this approach may overestimate the effect of GNPs as it does not account for absorption of energy within GNPs themselves and it is the enhanced dose to tissue (not gold) which is the quantity of interest for GNPT.  On the other hand, researchers have employed detailed simulations of discrete (not averaged or homogenized) GNPs within tissue/water considering microscopic volumes of interest \cite{Jo10,Le11,Zy13,KG15}.

Research to connect the macroscopic and microscopic approaches has been limited and varying.  To avoid modeling GNPs individually in a macroscopic volume yet still extract the DEF in pure tissue, Koger and Kirkby recently presented a database of factors to convert dose scored in a homogeneous blend of tissue and gold to dose scored in tissue containing discretely-modeled GNPs\cite{KK16}.  Some researchers have discretely modeled GNPs in a macroscopic volume using the simple geometry of a cubic lattice of GNPs embedded in an otherwise homogeneous water or tissue phantom  \cite{Me13,Gh12,Gh13,Zh09,To12b}.
%Another method of easing the computaional burden is to use a cubic lattice of GNPs, several studies \cite{Me13,Gh12,Gh13,Zh09,To12b} use a lattice (spanning $\sim1$~cm or slightly more \comment{Ghorbani has almost a factor of ten more at 9.42~cm$^3$, I don't know if slightly more does it justice}) embedded within an otherwise homogeneous water or tissue phantom, scoring dose to water or tissue in the lattice region.  
Cai \textit{et al}\cite{Ca13} used a lattice on a cellular scale, creating a multi-cell configuration containing discretely-modeled GNPs spanning (0.24~cm$)^3$.  Douglass \textit{et al}\cite{Do13} modeled single shapes (spheres or spherical shells) to represent aggregates of GNPs within a multi-cell model of total volume (0.04~cm$)^3$.  While the volumes considered in the latter two studies are above the microscopic scale, they are at least an order of magnitude too small for full GNPT simulations.

The current work focuses on bridging the gap between the microscopic and macroscopic paradigms.  We introduce % (for the first time, to our knowledge)
the \model (\acroc) model for MC simulations within the context of GNPT, a general framework based on the idea of combining distinct models of varying level of detail on different length scales in a single simulation.  In the current work, \acro models involve homogenized tissue-gold mixtures or pure tissue in larger ($\sim$cm) volumes and discrete modeling of GNPs embedded in tissue within distinct subvolumes.  %\comment{Gold in water was never mentionned here, but we do it for Sinha.  I honestly thinl it can be ommited as this sentance is already quite dense, just wanted to flag (just in case).}
%More conclusion - needed here?  The \acro model is a general framework, neither specific to one code nor to one geometric model or set of models.  It may be implemented within different MC radiation transport codes with advanced geometry packages to model diverse phantom dimensions while scoring within detailed microscopic models via efficient simulations.  
 %This is a general model, able to be implemented in any Monte Carlo code for radiation transport with an advanced geometry package.  It offers three distinct advantages over most models.  It can model virtually any size phantom, it can score dose in very detailed models and, perhaps most importantly, it is very efficient.  The results of this investigation achieve the same result as detailed single-scale model, but achieve the same statistical uncertainties one to two orders of magnitude faster.
%This computational approach of employing a homogenized, tissue-gold mixture within the bulk of a several centimeter phantom enables realistic representation of scatter conditions and determination of fluence within scoring volumes.  Within the scoring volumes, the discrete modeling of GNPs within multiple smaller ($\sim$\si{\micro\metre}) regions enables scoring of energy deposition within the tissue surrounding the GNPs.
The \acro approach is initially presented via example calculations with a cylindrical phantom, in which DEFs are computed at varying depths for photon sources of different energies.  A more sophisticated radiotherapy scenario recently presented\cite{Si15} is then considered, determining DEFs and comparing to published values computed with an analytic approach \cite{Ng10} used in many other studies.  Validation of simulations and considerations for future research are discussed.

% Parts of this paragraph were copy and pasted elsewhere ~~~~~~~~~~~~~~~~~~~~~~~~~~~~~~~~~~~~~~~~~~~~~~~~~~~~~~~~~~~~~~~~~~~~
%The \acro results demonstrate that the fluence perturbation due to GNPs and overestimation of dose scored in the homogeneous gold and tissue mixture are most substantial for lower source energies relevant for brachytherapy.  Some MC studies have investigated potential DEFs for brachytherapy involving GNPs\cite{To12,Zh09,Jo10}, with recent works proposing new avenues for GNP delivery\cite{Ci15,Si15}.  For accelerated partial breast irradiation, Cifter \textit{et al} propose using a polymer film embedded with GNPs coated on the balloon applicator for electronic brachytherapy \cite{Ci15}; the film dissolves over time in the patient, releasing GNPs which diffuse outward to create a significant concentration of gold in the treatment volume.  A similar approach is proposed by Sinha \textit{et al} for prostate permanent implant brachytherapy \cite{Si15} with GNPs released from inter-seed spacers throughout treatment.  For both works, DEFs are determined using a previously-developed analytical calculation approach\cite{Ng10}.

%  __  __      _   _               _     
% |  \/  |    | | | |             | |    
% | \  / | ___| |_| |__   ___   __| |___ 
% | |\/| |/ _ \ __| '_ \ / _ \ / _` / __|
% | |  | |  __/ |_| | | | (_) | (_| \__ \
% |_|  |_|\___|\__|_| |_|\___/ \__,_|___/
\section{Methods}

MC simulations are carried out using the EGSnrc\cite{Ka11} distribution with the egspp class library\cite{Ka05a}.  The user-code egs\_chamber\cite{Wu08} is used due to its variance reduction techniques, but is modified to enable energy-deposition scoring in multiple regions.  Transport parameters are generally EGSnrc defaults with the following exceptions: pair angular sampling is turned off, Rayleigh scattering and electron impact ionization are turned on, NRC cross section data are used for bremsstrahlung events and XCOM cross section data \cite{Be10} are used for photon interactions.  %Rather than employing the EGSnrc default of treating M- and N-shell transitions in an average way, 
Explicit $M$- and $N$-shell transitions are modeled to account for the dosimetric effects of these atomic relaxations\cite{WS16} (EGSnrc default is to treat $M$- and $N$-shell transitions in an average way); note that $K$- and $L$-shells are considered explicitly by default.  Photons and electrons are simulated down to 1 keV kinetic energy.  The high resolution random number generator option is enabled.  % to achieve a granularity much smaller than the scoring regions.
%The EADL\_RELAX macro is set to true, so as to use explicit (default is average) M and N-shell atomic transitions, required to accurately account for the effects of relaxation electrons generated in gold\cite{WS16}.

A cubic lattice geometry class is created to efficiently model GNPs within a medium discretely.  While cubic lattices have been employed in other studies\cite{Me13,Zh09}, the following describes our implementation within EGSnrc for the current work.  
%This geometry is used in several different studies in the literature\cite{Gh12,Me13,Zh09}; the following describes how it is implemented into EGSnrc for this study.  
The cubic lattice class takes a previously-defined geometry, chosen to be a gold sphere in the current work, and places it at $(an_x,an_y,an_z)$ for integers $n_x,n_y,n_z$ and an arbitrary spacing $a$ (which determines gold sphere number density); the same instance of the chosen geometry is placed at all possible positions within a volume.  This lattice geometry allows for large reductions in memory used, as well as reducing radiation transport times compared to a simulation in which each GNP geometry is modeled individually.  %RT - omit hexagonal lattice if we're not presenting the relevant calculations - it's irrelevant now. Using the code for the cubic lattice class, a hexagonal lattice is also created by adding four offset cubic lattices with shortened $x$-axis spacing to the same volume.  \comment{Rowan - do we need more details in last sentence?   But better to be minimalist} \comment{Maybe a quick reference to your paper to show what a hexagonal lattice looks like without getting into details?}
%The cubic lattice geometry allows for large reductions in memory used, as well as having fast electron and relatively fast photon transport when comparing to a simulation where each lattice geometry is modelled individually.  The efficiency of the photon transport scales linearly with the spacing between lattice geometries, the mean path length of the travelling photon and the complexity of the lattice geometry.  RT - keep details for later work; not needed for Med Phys letter.
In simulations where the cubic lattice is irradiated by a parallel beam, the lattice is tilted (15$^\circ$ about one perpendicular axis and 30$^\circ$ about the other) relative to the beam axis to avoid shadowing (\ie having all primary photon trajectories either intersecting with all the GNPs in a row or none at all) which would create a bias that would not exist in a random distribution.  Results in this study are not sensitive to the angles chosen as long as the cubic lattice axes are not close to parallel (within 3$^\circ$) with the beam.  %RT suggest omitting next part of sentence because wishy-washy:", though the sensitivity would vary in different scenarios."

%it is noted that the 3 axes of the lattice should differ from the beam axis to avoid shadowing, \ie by tilting the lattice axes relative to the beam axis, we avoid having all primary photon trajectories either intersecting with all the GNPs in a row or none at all, creating a bias that would not exist in a random distribution.

\subsection{Verification of MC simulations with GNPs}

We perform a verification of our MC simulations involving GNPs using the recent (independent) results of Koger and Kirkby\cite{KK16}.  Using PENELOPE, these researchers modeled monoenergetic photons or electrons incident on microcavities filled with either randomly-distributed GNPs embedded in ICRU tissue\cite{ICRU44} or a homogeneous mixture of gold and tissue.  Photons and electrons were simulated down to 100~eV, and electrons leaving the microcavity were transported back into it for full energy deposition.  Ratios of the dose-to-tissue relative to dose to the homogeneous blend of tissue and gold were published (to allow conversion of dose scored in a mixture to dose-to-tissue) for a range of source energies and GNP concentrations\cite{KK16}.  

\begin{figure}[htbp]
	\centering
	\includegraphics[width=0.45\textwidth]{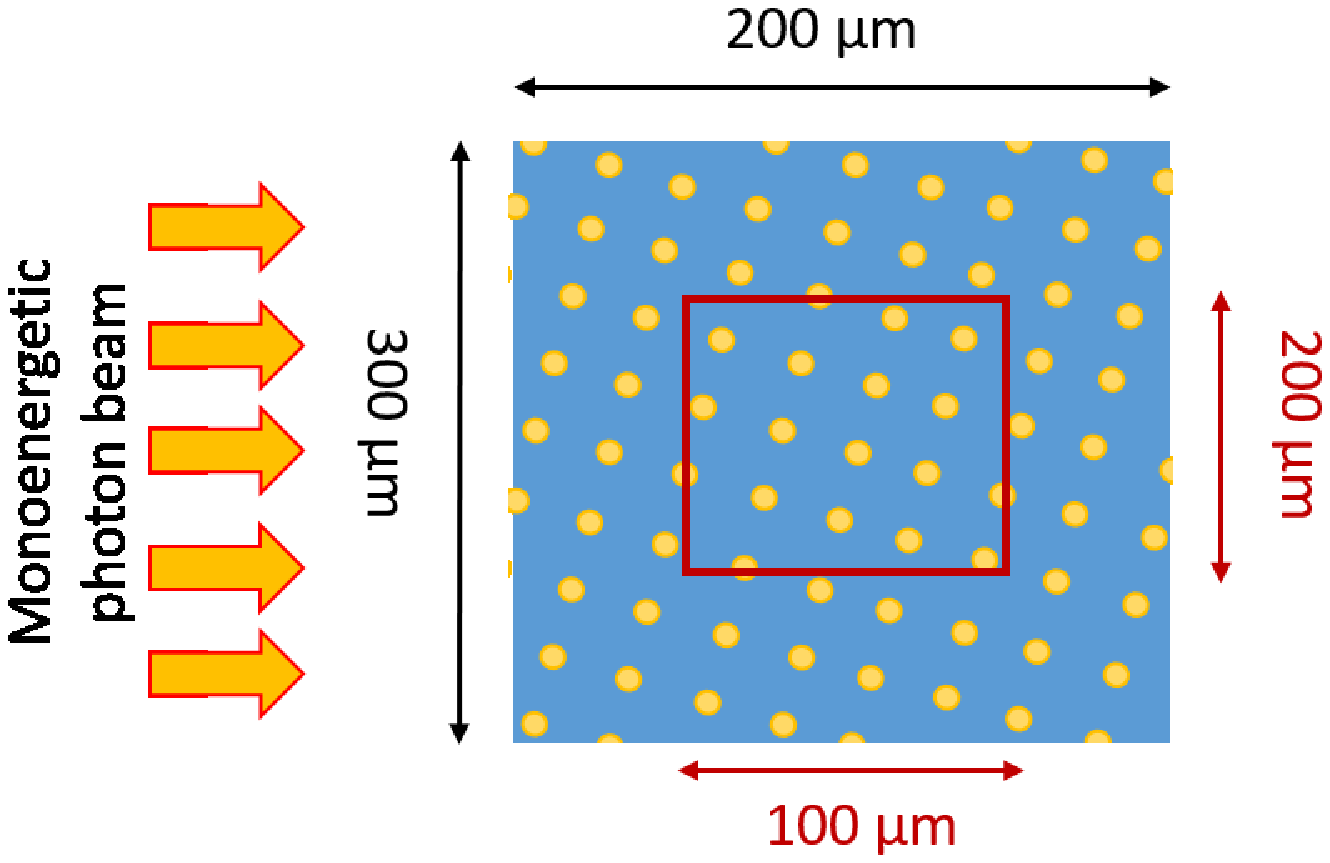}
	\captionl{Schematic diagram for verification of MC simulations with GNPs: cross section of the cylindrical phantom comprised of a lattice of GNPs embedded in tissue; the scoring region, outlined in red, is at the center of the larger phantom.  For the homogeneous scenario, the cylindrical phantom consists of a single medium that is a homogeneous mixture of gold and tissue. 
	\label{fig:MicroCyl}}
\end{figure}

In the current work, dose ratios (dose-to-tissue relative to dose-to-mixture of tissue and gold) are calculated with simulations involving cylindrical geometries.  A monoenergetic (20, 30 or 50~keV) parallel photon beam (circular cross section, 150~\si{\micro\metre} radius) is incident on a cylinder (150~\si{\micro\metre} radius, 200~\si{\micro\metre} long) containing either a lattice of GNPs (20 or 100~nm diameter) embedded in pure ICRU tissue \cite{ICRU44}  or a homogeneous mixture of gold and ICRU tissue (concentrations of 5, 10 and 20~mg/g).  Dose is scored in a smaller cylinder (100~\si{\micro\metre} radius, 100~\si{\micro\metre} long) located at the center of the larger cylinder (see Fig.~\ref{fig:MicroCyl}\mfig{fig:MicroCyl}).  %, to achieve scatter conditions.
% so as to achieve forward and lateral electron build up in the scoring region.  
The dose ratio (or conversion factor) is computed by taking the ratio of the doses scored in the GNP lattice to homogeneous scenarios.  %Comparisons are only made for conversion factors deviating from unity and hence no simulations for sources with energies above 50 keV are performed.

\subsection{Parallel beam of monoenergetic photons incident on cylinder}\label{meth:cylinder}

\begin{figure}[htbp]
	\centering
	\includegraphics[width=0.48\textwidth]{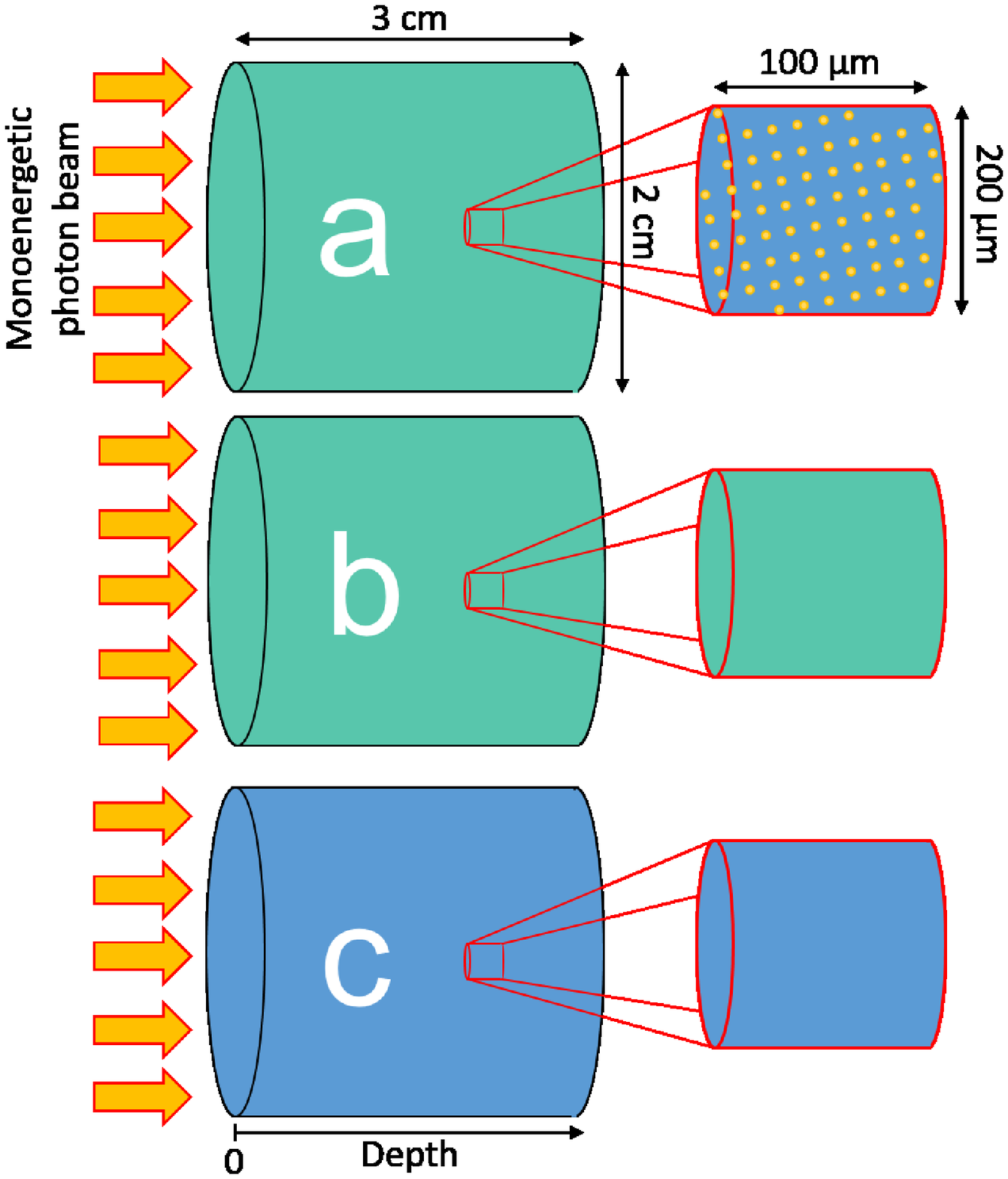}
	\captionl{Schematic diagrams for cylindrical phantom simulations: (a) \acro model with homogeneous gold-tissue phantom with 119 smaller cylinders consisting of GNPs embedded in tissue placed at even intervals along the central axis; (b) corresponding fully homogeneous (gold-tissue mixture) phantom; and (c) pure tissue phantom.
	%Cylindrical phantom geometries to illustrate the heterogeneous multiscale (HM) model.  
	%Diagram shows the scatter phantom for the three different cases used to determine DEF versus depth on the left, and a sample scoring region for each case located along the central axis to the right (not to scale).  Cylinder a and b both use a homogeneous mixture of gold and tissue for the scatter phantom, but have different scoring regions.  Cylinder b models a homogeneous scoring region as well, where as a models the gold discretely in a lattice (GNPs also not to scale).  Cylinder c shows the pure tissue case.	
	\label{fig:DiagramCylinders}}
\end{figure}

For the first set of calculations using the \acro model, parallel beams (circular cross section, 1~cm radius) of monoenergetic photons with energies 20, 30, 50, 90, and 200~keV, as well as 1~MeV are incident on cylindrical phantoms (1~cm radius, 3~cm long), depicted in Fig.~\ref{fig:DiagramCylinders}\mfig{fig:DiagramCylinders}.  In the \acro model, the cylinder is comprised of a homogeneous mixture of ICRU tissue\cite{ICRU44} and gold, containing smaller cylinders (100~\si{\micro\metre} radius, 100~\si{\micro\metre} long) comprised of a lattice of discretely-modeled spherical GNPs within (pure) ICRU tissue (Fig.~\ref{fig:DiagramCylinders}a).  A total of 119 smaller cylinders evenly-spaced along the central axis of the cylindrical phantom are simultaneously modeled and energy deposited within the tissue (not GNPs) is scored.   For comparison, a fully homogeneous cylinder is modeled in which all regions are a mixture of tissue and gold and dose is scored therein (Fig.~\ref{fig:DiagramCylinders}b).  Simulations are then repeated with a tissue-only phantom to provide the denominators for computation of DEFs (Fig.~\ref{fig:DiagramCylinders}c).  
Simulations are performed for concentrations of 5, 10 and 20~mg/g, and GNP diameters of 20 and 100~nm.  Results for the \acro model (Fig.~\ref{fig:DiagramCylinders}a) are compared to those from simulation of the cylinder comprised of tissue entirely filled with a GNP lattice (for a subset of source energies and gold concentrations) for further validation, as well as characterization of simulation efficiencies.

% Cell dimensions
% 5E-4 0.000501076 7.35E-4
% 5E-4 0.000500273 7.35E-4

%An additional \acro model (Fig.~\ref{fig:DiagramShells}\mbox{fig:DiagramShells}) is considered motivated by the fact that GNPs may be clustered within biological structures: a spherical shell of gold about the cell nucleus is modelled\cite{Do13} representing the ideal GNP aggregation about the nuclear membrane\cite{Li08d}.  Following the cell configuration of Thomson {\it et al}\cite{Th13}, cells are placed in a hexagonal lattice, with a nuclear radius of 5~\si{\micro\metre} and a total cell radius of 7.35~\si{\micro\metre}, with a total cell density of $\num{3E8}$ cells per cubic centimetre.  For ease of comparison with the other cylindrical phantom simulations, all non-gold regions are assigned tissue (rather than cell cytoplasm, nucleus, extracellular matrix elemental compositions \cite{Th13}).  Dose to the nucleus (which is the region surrounded by the spherical shell of gold), thus computing nuclear DEFs.  We consider incident photon energies of 20 and 50~keV, and concentrations of 5 mg/g (2.73~nm thick shells) and 20 mg/g (10.76~nm thick gold shells).  

%\begin{figure}[htbp]
%	\centering
%	\includegraphics[width=0.95\textwidth]{Martinov_Figure3.eps}
%	\captionl{Schematic diagram of a more complex cylindrical phantom simulation using the same scatter and scoring cylinder dimensions as in Fig.~\ref{fig:DiagramCylinders}, but with different GNP arrangement.  The GNPs are simulated as spherical shells about the nucleus of cells arranged in a hexagonal lattice with a number density of $3\times10^8$.  Cells are modeled as two concentric spheres, 5 and 7.35~\si{\micro\metre} radius spheres representing the nucleus and cytoplasm, respectively.  The thickness of the gold shell is such that the total concentration of gold in the scoring region does not change, and all non-gold media are still ICRU tissue.
%	\label{fig:DiagramShells}}
%\end{figure}

\subsection{Brachytherapy source in a sphere with varying gold concentration}

%Some MC studies have investigated potential DEFs for brachytherapy involving GNPs\cite{To12,Zh09,Jo10}, with recent works proposing new avenues for GNP delivery\cite{Ci15,Si15}, due to the large photoelectric cross section of gold at brachytherapy source energies.  

The second scenario modeled in the \acro framework is from the work of Sinha \textit{et al}\cite{Si15} who considered a polymer film embedded with GNPs coated on the inter-seed spacers used in prostate permanent implant brachytherapy; the film dissolves over time in the patient, releasing GNPs which diffuse outward to create a significant concentration of gold in the treatment volume.  Sinha \textit{et al} calculated DEFs using a previously-developed analytical calculation approach\cite{Ng10}: calculations were performed at different radii within a water sphere containing an isotropic point source ($^{125}$I, $^{103}$Pd and $^{131}$Cs) at its center.  Assuming GNPs steadily released from the sphere center and, using Fick's second law of diffusion\cite{Cr79}, GNP concentration was determined as a function of time and radial position within the sphere, enabling computation of DEFs.  No particular localization of GNPs within tumor cells was assumed, \ie GNPs randomly arranged within a region of a given concentration.

\begin{figure}[htbp]
	\centering
		\begin{subfigure}{0.45\textwidth}
			\centering
			\includegraphics[width=0.99\textwidth]{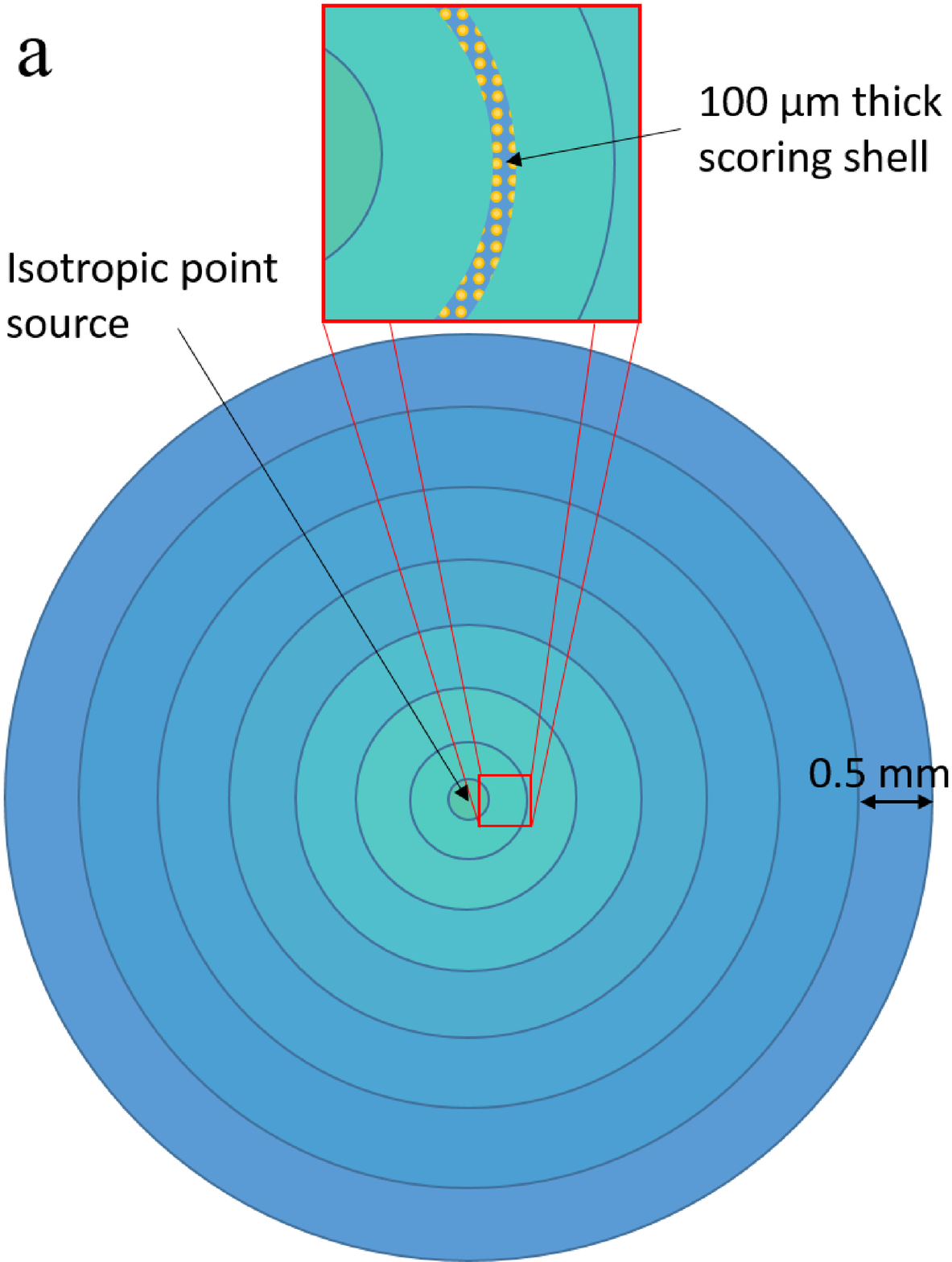}
		\end{subfigure}
		\begin{subfigure}{0.45\textwidth}
			\centering
			\includegraphics[width=0.99\textwidth]{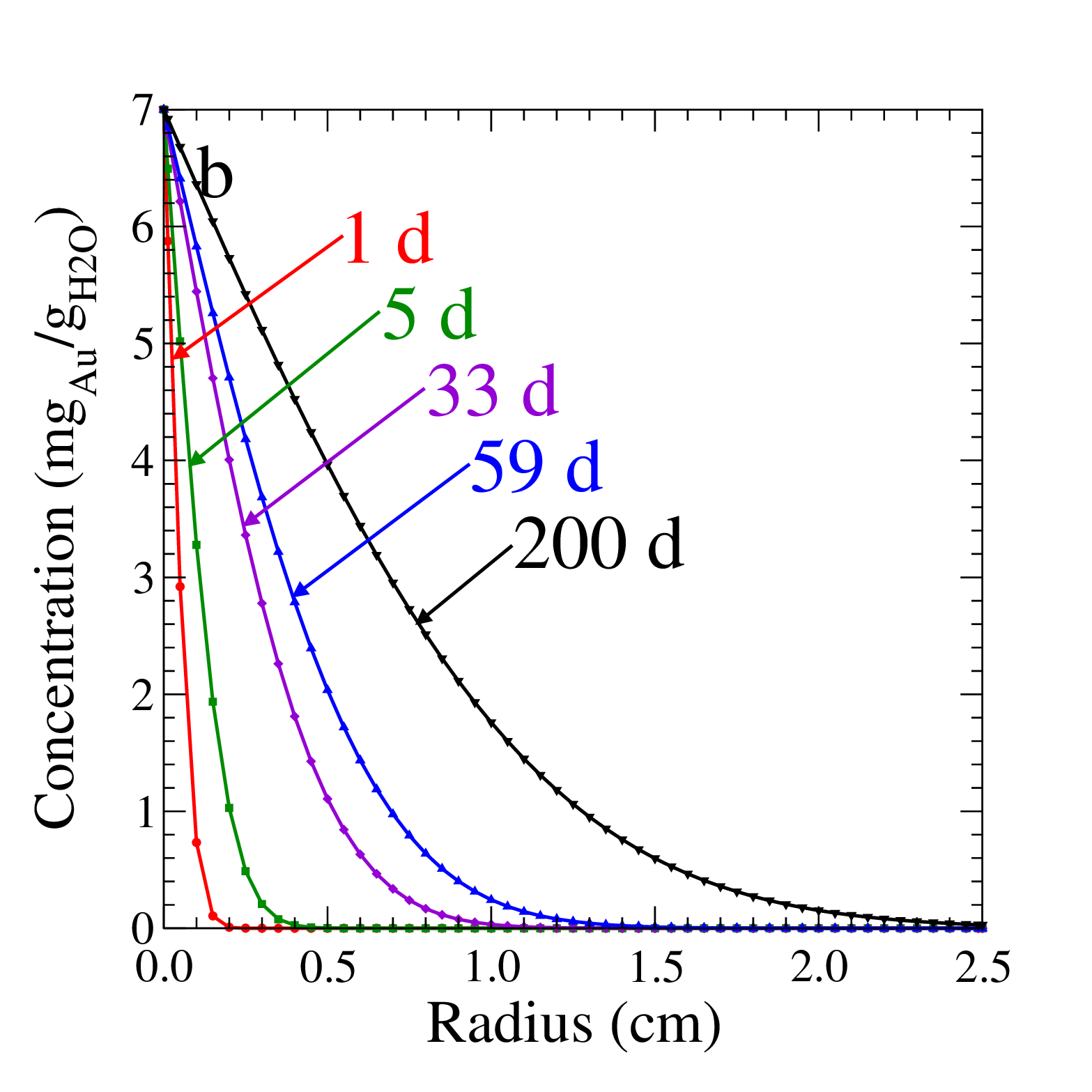}
		\end{subfigure}
	\captionl{(a) Schematic cross-sectional diagram of the sphere with GNPs released from its center; homogeneous water-gold mixtures are modeled in spherical shells 0.5~mm thick.  A 100~\si{\micro\metre} thick spherical shell consisting of discretely-modeled GNPs in water (shown above) is embedded in each shell.  (b) Gold concentrations within the sphere as a function of radius for different times after implantation assuming 7~mg Au/g water at the center and following Fick's second law of diffusion. \label{fig:DiagramSpheres}}
\end{figure}

In the \acro model simulation of the above scenario, a 2.5~cm radius sphere divided into 0.5~mm shells each containing a homogeneous mixture of gold and water with concentrations based on the diffusion of GNPs from the center is modeled (Fig.~\ref{fig:DiagramSpheres}\mfig{fig:DiagramSpheres}).  The 50 shells are a discrete approximation of the continuous decrease in concentration expected.  At the center of each homogeneous shell, a 100~\si{\micro\metre} thick spherical shell comprised of 20~nm diameter GNPs in a lattice embedded in water is modeled, and dose is scored in water within these regions.  An isotropic point source at the center of the sphere is used assuming the spectra leaving the following seed models\cite{Ch16}: OncoSeed 6711\cite{Do06,NCRP58}~($^{125}$I), IsoRay CS1\cite{Mu04,NNDC}~($^{131}$Cs), and  TheraSeed 200\cite{MW02,NNDC}~($^{103}$Pd).  Simulations are repeated with an all-water geometry (no gold) to provide the denominators to compute DEFs.

%  _____                 _ _       
% |  __ \               | | |      
% | |__) |___  ___ _   _| | |_ ___ 
% |  _  // _ \/ __| | | | | __/ __|
% | | \ \  __/\__ \ |_| | | |_\__ \
% |_|  \_\___||___/\__,_|_|\__|___/
\section{Results}
\subsection{Verification of MC simulations with GNPs}\label{res:KK16}

Table~\ref{tab:Koger}\mtab{tab:Koger} presents dose ratios (dose-to-tissue relative to dose to a homogeneous tissue-gold mixture) for various gold concentrations and source energies.  %RT- stated in methods so omit here: at which the dose is sensitive to the model of gold in the scoring region.  
While many values computed in the current work agree with the published values of Koger and Kirkby \cite{KK16} within the 1$\sigma$ statistical uncertainties indicated, all values agree within 2$\sigma$ uncertainties except for the 20 keV beam with 20~nm diameter GNPs at a concentration of 20~mg/g, which is at 2.1$\sigma$. %2.117
The largest percent difference observed is 2.6\% for the 50~keV beam with 20~nm diameter GNPs having a concentration of 20~mg/g (quadrature sum of the 1$\sigma$ statistical uncertainties is 1.3\%).  The absolute difference between our results and those of Koger and Kirkby averaged over all the source energies, GNP diameters, and gold concentrations considered is 0.92\%.  Simulations repeated with larger cylindrical phantoms (radii of 200 or 300~\si{\micro\metre} rather than 150~\si{\micro\metre}, lengths of 300 or 400~\si{\micro\metre} rather than 200~\si{\micro\metre}) but with the same central scoring volume (radius 100~\si{\micro\metre}, length 100~\si{\micro\metre}) yield dose ratios in agreement within statistical uncertainties.
  %\comment{I'm gonna redo the 100 and 200 micron buildup ones making sure its padded on all sides with 100 or 200 microns.}%\comment{I looked at TG195 and saw basically 0.2\% between PENELOPE and EGS, but I dug up one of Ernesto's papers where he did a comparison of codes, and eh found that for an electron beam incident on a sphere of water sitting in a vacuum, EGS scored systematically less energy deposition than PENELOPE.  I've added the paper to the zip file if you want to see it.}

%Hi Rowan,
%    On a plot they would overlap, but as far as I've been taught, agreement states how many standard
%deviations from zero the distribution of the difference is (the t-test).  If you add error in quadrature,
%then it is above 2 standard deviations.  If adding in quadrature, then only 9 points are within 1 sigma, 8
%points within 2 sigma and 1 point above 2 sigma.  If just look at uncertainty linearly, then you are right 12
%within 1 sigma and 6 within 2 sigma.
%Martin

\begin{table}[htbp]
\vspace{3mm}
	\captionl{Verification of Monte Carlo simulations: Comparison of dose ratios (dose-to-tissue relative to dose to a homogeneous tissue-gold mixture) for 20 and 100~nm diameter GNPs for the current work (Fig.~\ref{fig:MicroCyl})  and the independent results of Koger and Kirkby \cite{KK16}.  Statistical uncertainty (1$\sigma$) on the final digit(s) is indicated in parentheses.  \label{tab:Koger}}
	\begin{tabular}{ccllllll}
		\hline\hline
		GNP &&& \multicolumn{2}{l}{20~nm diam. GNPs} && \multicolumn{2}{l}{100~nm diam. GNPs} \\
		\cline{4-5} \cline{7-8}
		%\parbox[c][1.2cm]{1.6cm}{Concen.\\(mg$_{\text{Au}}$/g$_{\text{tissue}}$)} & \parbox[c][1.2cm]{1.6cm}{Energy\\(keV)} & \parbox[c][1.2cm]{2cm}{Koger and\\Kirkby} & \parbox[c][1.2cm]{2cm}{This\\work} && \parbox[c][1.2cm]{2cm}{Koger and\\Kirkby} & \parbox[c][1.2cm]{2cm}{This\\work} \\
		Concentration & Energy && Koger and & This && Koger and & This \\
		(mg$_{\text{Au}}$/g$_{\text{tissue}}$) & (keV) && Kirkby & work && Kirkby & work \\
		\hline
		   & 20 && 0.949 (6)  & 0.951 (3) && 0.884 (7)  & 0.882 (3) \\
		 5 & 30 && 0.969 (8)  & 0.963 (5) && 0.927 (7)  & 0.930 (5) \\ \vspace{2mm}
		   & 50 && 0.984 (12) & 0.990 (9) && 0.974 (10) & 0.950 (9) \\ 
		   & 20 && 0.937 (6)  & 0.926 (2) && 0.827 (6)  & 0.824 (2) \\
		10 & 30 && 0.950 (8)  & 0.947 (4) && 0.910 (7)  & 0.901 (4) \\ \vspace{2mm}
		   & 50 && 0.984 (11) & 0.968 (8) && 0.946 (10) & 0.952 (7) \\ 
		   & 20 && 0.917 (7)  & 0.902 (2) && 0.791 (6)  & 0.782 (2) \\
		20 & 30 && 0.952 (8)  & 0.939 (3) && 0.881 (7)  & 0.875 (3) \\
		   & 50 && 0.986 (11) & 0.961 (6) && 0.942 (10) & 0.929 (6) \\
		\hline\hline\vspace{2mm}
	\end{tabular}
\end{table}

\subsection{Parallel beam of monoenergetic photons incident on cylinder}\label{res:cylinder}

\begin{figure}
	\centering
		\begin{subfigure}{0.49\textwidth}
			\centering
			\includegraphics[width=0.99\textwidth]{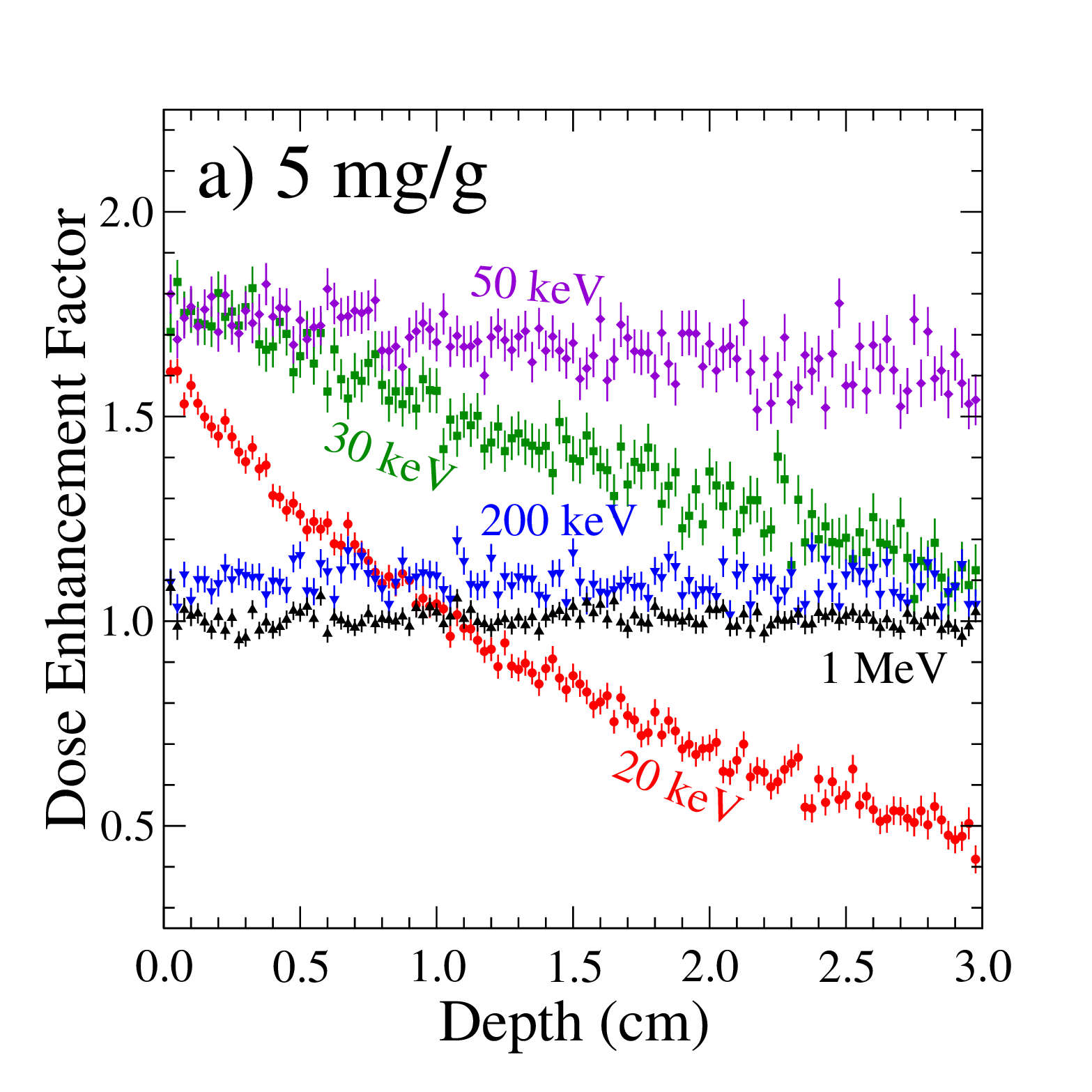}
		\end{subfigure}
		\begin{subfigure}{0.49\textwidth}
			\centering
			\includegraphics[width=0.99\textwidth]{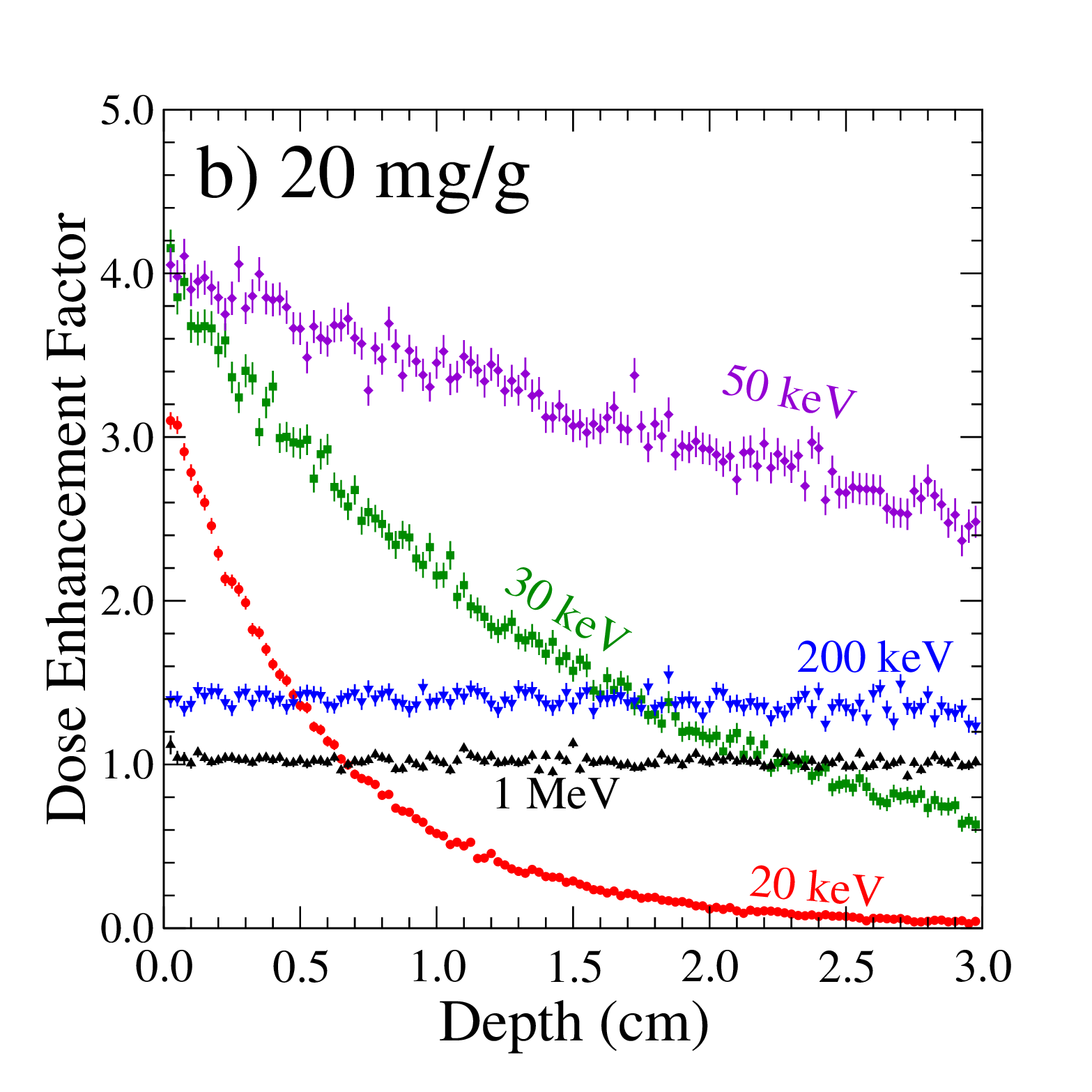}
		\end{subfigure}
	\captionl{DEFs versus depth within the cylindrical phantom (Fig.~\ref{fig:DiagramCylinders}) with concentrations of (a) 5 and (b) 20~mg of Au per g of tissue for 20~nm diameter GNPs with incident photon source energies of 20~keV to~1 MeV.  DEFs for 90~keV which show smaller magnitude but similar trends to the results for 50~keV are omitted for clarity.  Results for simulations with a concentration of 10~mg/g  (omitted) exhibit similar trends to 5 and 20~mg/g.  \label{fig:PDD}}%  (d) DEFs for 100~nm GNPs compared to those for 20~nm GNPs for the 5~mg/g.  In each panel, DEFs computed within the \model (\acroc) and homogeneous models are shown.  Error bars indicate statistical uncertainties.  \label{fig:PDD}}
\end{figure}

The results of the cylinder simulations for 20~nm diameter GNPs are summarized in Figure~\ref{fig:PDD}\mfig{fig:PDD}.  Focusing first on photon source energies below 200~keV, DEFs for all concentrations are most substantial near the surface of the phantom, ranging from 1.6 to 4.2, and decrease with depth.  DEFs decrease more rapidly with increasing depth in the 20~mg/g case: for the 50~keV source, DEFs decrease from 4 to 2.5 over the 3~cm length of the phantom, while DEFs for the 20~keV source are near 3 at the surface and decrease nearly to zero.  DEFs decrease below unity for some gold concentrations and source energies: for 5~mg/g and a 20~keV source, there are DEFs$<1$ for depths of 1~cm or more; for 10 and 20~mg/g, DEFs drop below unity within the 3 cm long phantom for a 30~keV beam as well.   %Dose enhancements result from enhanced local energy deposition due to photoelectric interactions in GNP; dose gradients arise due to decreasing fluence with depth due to the presence of gold between the front of the cylinder and scoring depth. 
These phenomena arise due to competing effects of enhanced local energy deposition (due to photoelectric interactions in local GNP) and decreasing fluence with depth (due to the presence of gold between the front of the cylinder and the scoring depth, \ie within the scatter media), and are expected on the basis of energy conservation.  % Move Robar reference to lower down because there are relevant GNPT works that also have results with gradient (whether they comment on it or not) Within the context of tumor dose enhancement using modified megavoltage beams and gadolinium or iodine contrast media (not GNPT), Robar {\it et al} noted analogous changes in dose enhancement over centimeter length scales \cite{Ro12a}.
%These effects are noted previously in other work scoring DEF on cm length scales\cite{Ro12a}, but isolate them more in the \acro model case by scoring only in tissue.  
When the effects of the decreasing fluence exceed local energy deposition gains, dose scored within the \acro model is lower than that in the pure tissue phantom (no gold), resulting in DEFs below unity, \ie dose \textit{decreases} not \textit{enhancements}, making DEF a misnomer.  Despite this, the dose ratio will continue to be referred to as a DEF for the purposes of continuity.

As source energy increases, the relative importance of photoelectric events decreases resulting in DEFs nearer unity and less variation in DEFs with depth.  For 200~keV photons, DEFs are approximately constant with depth within the 3~cm phantom, at a value near 1.05 for 5~mg/g and 1.40 for 20~mg/g.   For the 1 MeV beam, incoherent (Compton) interactions dominate resulting in DEFs near unity for the length of the cylinder -- no significant dose enhancement.

\begin{figure}[htbp]
	\centering
		\begin{subfigure}{0.49\textwidth}
			\centering
			\includegraphics[width=0.99\textwidth]{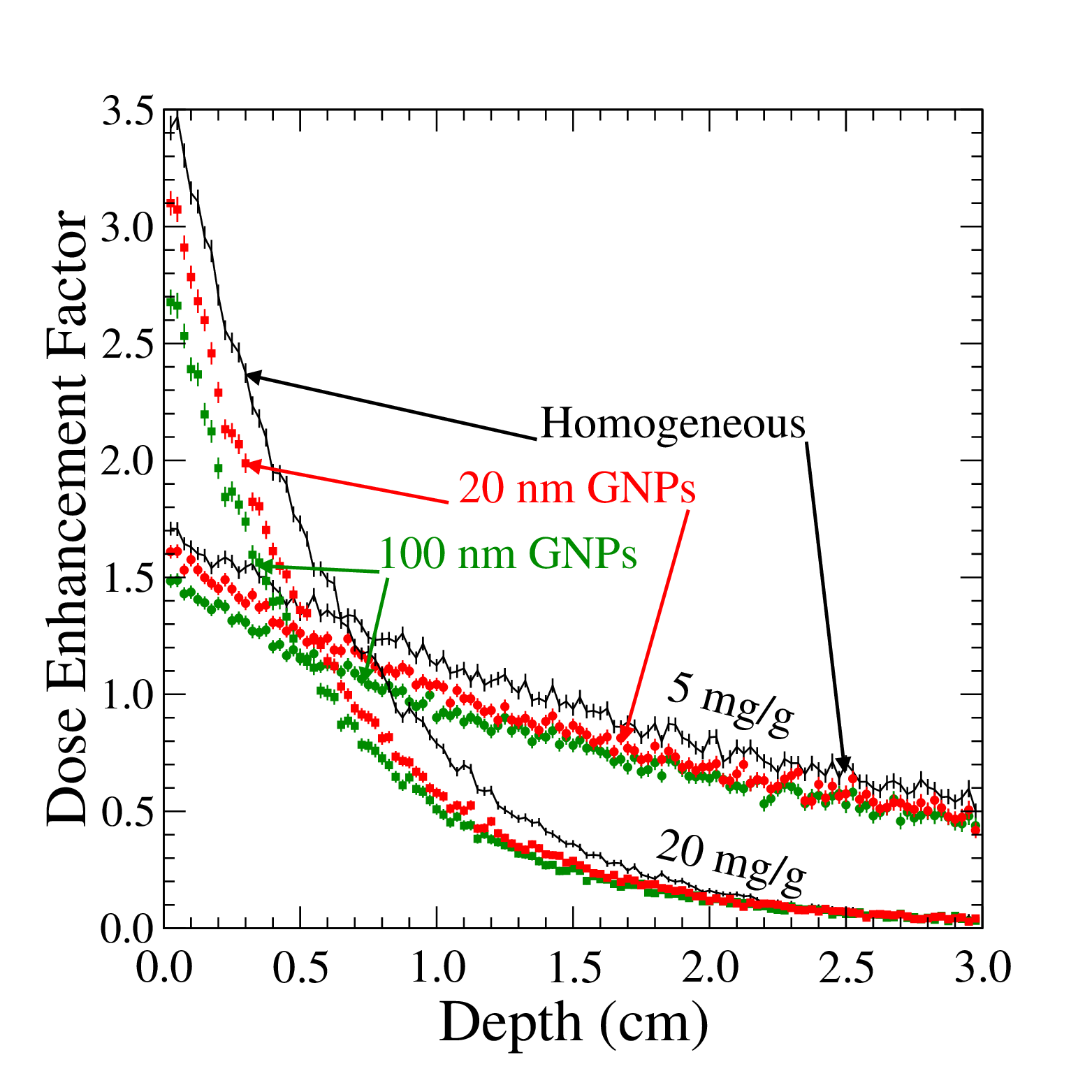}
		\end{subfigure}
	\captionl{DEFs versus depth within the cylindrical phantom (Fig.~\ref{fig:DiagramCylinders}) with concentrations of 5 and 20~mg/g for 20 and 100~nm diameter GNPs with 20~keV incident photons.  DEFs computed for the 30 and 50~keV source energies show similar trends to the 20~keV case but are omitted for clarity.  \label{fig:PDD_rad}}
\end{figure}

DEFs computed with 20 or 100~nm diameter GNPs in scoring regions (\acro model) along with the homogenized tissue-gold mixture are summarized in Figure~\ref{fig:PDD_rad}\mfig{fig:PDD_rad} for the 20~keV photon source.    In general, DEFs within the homogeneous model (qualitatively) follow the same trends as those in the \acro model, showing competition between attenuation of fluence and enhanced energy deposition.  However, for all gold concentrations and GNP diameters, DEFs computed within the homogeneous model overestimate those for the \acro model: discrepancies are as large as 20\% for the 20~keV beam and vary with depth.  As source energy increases, discrepancies between DEFs computed with the \acro and homogenized models decrease (results not shown; see also dose ratios presented in table~\ref{tab:Koger}, section~\ref{res:KK16}).  The overestimation observed for DEFs computed in the homogenized model stems from the inclusion of energy deposited within gold into the total `enhanced' dose, while it is only dose to tissue (not gold) that is relevant for GNPT\cite{Le11,Gh12}.  In the \acro simulations with GNPs modeled discretely and dose in tissue scored, photoelectrons generated in the gold may deposit all their energy in the gold or they must travel some distance (and deposit energy) before entering tissue.  By energy conservation, there is then less dose in the numerator for the computation of DEFs.  Comparing \acro model results for 20 and 100~nm diameter GNPs but same total gold concentration, DEFs are consistently higher with smaller GNPs for lower source energies (20 and 30~keV) but do not vary significantly at higher energies.  Sensitivity to GNP size stems from the re-absorption of electrons generated within GNPs which is more important for larger GNPs \cite{Le11,Gh12}. %Already cited these earlier in paragraph so omit here:  (see, \eg  Lechtman {\it et al}\cite{Le11}, Ghorbani {\it et al} \cite{Gh12}).

All results presented with the \acro model thus far might have been carried out in a phantom with each GNP modeled discretely, rather than combining models with different features on different length scales, as done for the \acro model.  We carried out a subset of simulations in a cylindrical phantom comprised of ICRU tissue with a lattice of GNPs spanning the entire phantom, and results agreed with those computed within the \acro model with GNPs only modeled in subvolumes (Fig.~\ref{fig:DiagramCylinders}) within statistical uncertainties.  %We may be asked to fill in what those stat unrest are but we'll leave as is for now because uncertainties vary so much in different parts of the phantom etc.
The major difference between the two sets of simulations was time or, equivalently, efficiency:
\begin{equation}
\epsilon = \frac{1}{s^2 t}\, , \label{eq:eff}
\end{equation}
where $t$ is the simulation time and $s$ is the average percent uncertainty on doses in the 119 scoring regions (spanning the 3~cm long cylindrical phantom).  Table~\ref{tab:eff} demonstrates considerable efficiency gains with the \acro model compared with having a lattice filling the entire volume of interest, with efficiencies enhanced by factors of 24 (20~mg/g; 20~keV) to 122 (5~mg/g; 50~keV); \acro simulations have lower efficiencies than the corresponding (inaccurate) homogeneous models by 15\% (5~mg/g; 20~keV) to 31\% (20~mg/g; 20~keV).  Efficiencies vary with source energy and gold concentration, \eg the lower efficiency of the 20 keV photons and gold concentration 20~mg/g reflects larger statistical uncertainties at depth in the cylinder (due to considerable attenuation of these relatively low energy photons with the higher concentration of gold).

Simulation times (on a single Intel Xeon 5160 core) may be deduced from the results in Table~\ref{tab:eff}: for the concentrations and source energies considered therein, \acro simulation times for $s=2\%$ average uncertainty range from 0.12~years (5~mg/g; 50~keV) to 0.91~years (20~mg/g; 20~keV) while full lattice simulation times range from 3.9~years (5~mg/g; 20~keV) to 22~years (20~mg/g; 20~keV).  The results presented in Figs.~\ref{fig:PDD} and \ref{fig:PDD_rad} are from simulations of $>10^{10}$ histories, with number of histories adjusted to achieve the relatively small error bars (depending on source energy and gold concentration).  We have not included the times for these particular calculations because simulations were run on multi-core clusters with CPUs of varying speeds, and hence these calculation times (or efficiencies) are not representative measures.

\begin{table}[htbp]
\vspace{3mm}
	\captionl{Example efficiencies (Eq.~\ref{eq:eff}) for \acro simulations in the cylindrical phantom (Fig.~\ref{fig:DiagramCylinders}a), with efficiencies relative to the tissue cylinder entirely filled with a lattice (``HetMS/Lattice'') and to the homogeneous gold-tissue mixture cylinder (Fig.~\ref{fig:DiagramCylinders}b; ``HetMS/Homog'').  Estimates of efficiencies are based on calculation times for simulation of $3.6\times10^6$ histories on a single thread Intel Xeon 5160 (3.00~GHz) core.  The time to achieve $s=$1\% is $1/\epsilon$. %Note that only the lattice and HetMS simulations provide accurate results, the discrepancies of the homogeneous model are discussed in \ref{res:cylinder}.
	\label{tab:eff}}
	\begin{tabular}{cccccc}
		\hline\hline
		Concentration & Energy & Efficiency  & \multicolumn{2}{c}{Relative efficiency}\\ \cline{4-5} 
		(mg$_{\text{Au}}$/g$_{\text{tissue}}$)& (keV) & HetMS (h$^{-1}$) & HetMS/Lattice & HetMS/Homog \\
		\hline
		 5 & 20 & $2.2 \times 10^{-4}$ &  29 & 0.85 \\ 
		   & 50 & $2.4 \times 10^{-4}$ & 122 & 0.69 \vspace{2mm}\\ 
		20 & 20 & $3.1 \times 10^{-5}$ &  24 & 0.73	\\ 
		   & 50 & $2.1 \times 10^{-4}$ &  89 & 0.72 \\ 
		\hline\hline\vspace{2mm}
	\end{tabular}
\end{table}

%\begin{table}[htbp]
%\vspace{3mm}
%	\captionl{Example efficiencies ($\frac{1}{s^2\times{}t}\times1000$ where $s$ is the square root quadrature sum of all the uncertainties and $t$ is the computation time), with relative efficiencies to the \acro model, for simulations of $3.6\times10^6$ histories on a single CPU for a full lattice scenario, the HetMS model (Fig.~\ref{fig:DiagramCylinders}a) and a full homogeneous scenario (Fig.~\ref{fig:DiagramCylinders}b) on the same machine.  Note that only the lattice and HetMS simulations provide accurate results, the discrepancies of the homogeneous model are discussed in \ref{res:cylinder}.
%	\label{tab:eff}}
%	\begin{tabular}{ccccc}
%		\hline\hline
%		Concentration & Energy &  \multicolumn{3}{c}{Efficiency (s$^-1$)}\\ \cline{3-5} 
%		(mg$_{\text{Au}}$/g$_{\text{tissue}}$)& (keV) & Lattice & HetMS & Homogeneous \\
%		\hline
%		 5 & 20 & 0.6 (0.008) &  79 & 104 (1.447) \\ \vspace{2mm}
%		   & 50 & 5.0 (0.034) & 150 & 168 (1.175) \\ 
%		20 & 20 & 4.8 (0.042) & 116 & 153 (1.368) \\ \vspace{2mm}
%		   & 50 & 0.8 (0.011) &  89 &  97 (1.390) \\ 
%		\hline\hline\vspace{2mm}
%	\end{tabular}
%\end{table}

\subsection{Brachytherapy source in a sphere with varying gold concentration}\label{res:sphere}

\begin{figure}[htbp]
	\centering
		\begin{subfigure}{0.49\textwidth}
			\centering
			\includegraphics[width=0.99\textwidth]{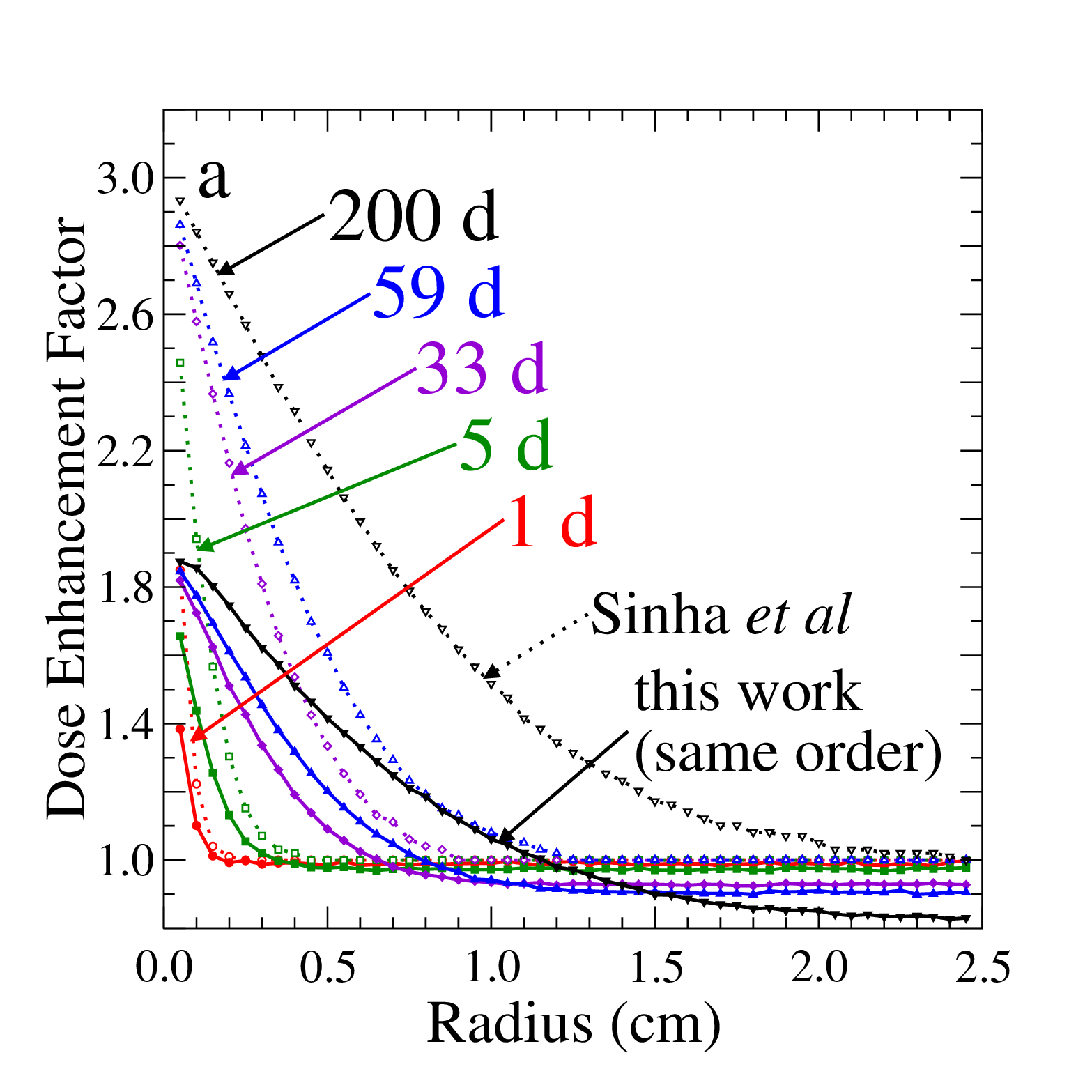}
		\end{subfigure}
		\begin{subfigure}{0.49\textwidth}
			\centering
			\includegraphics[width=0.99\textwidth]{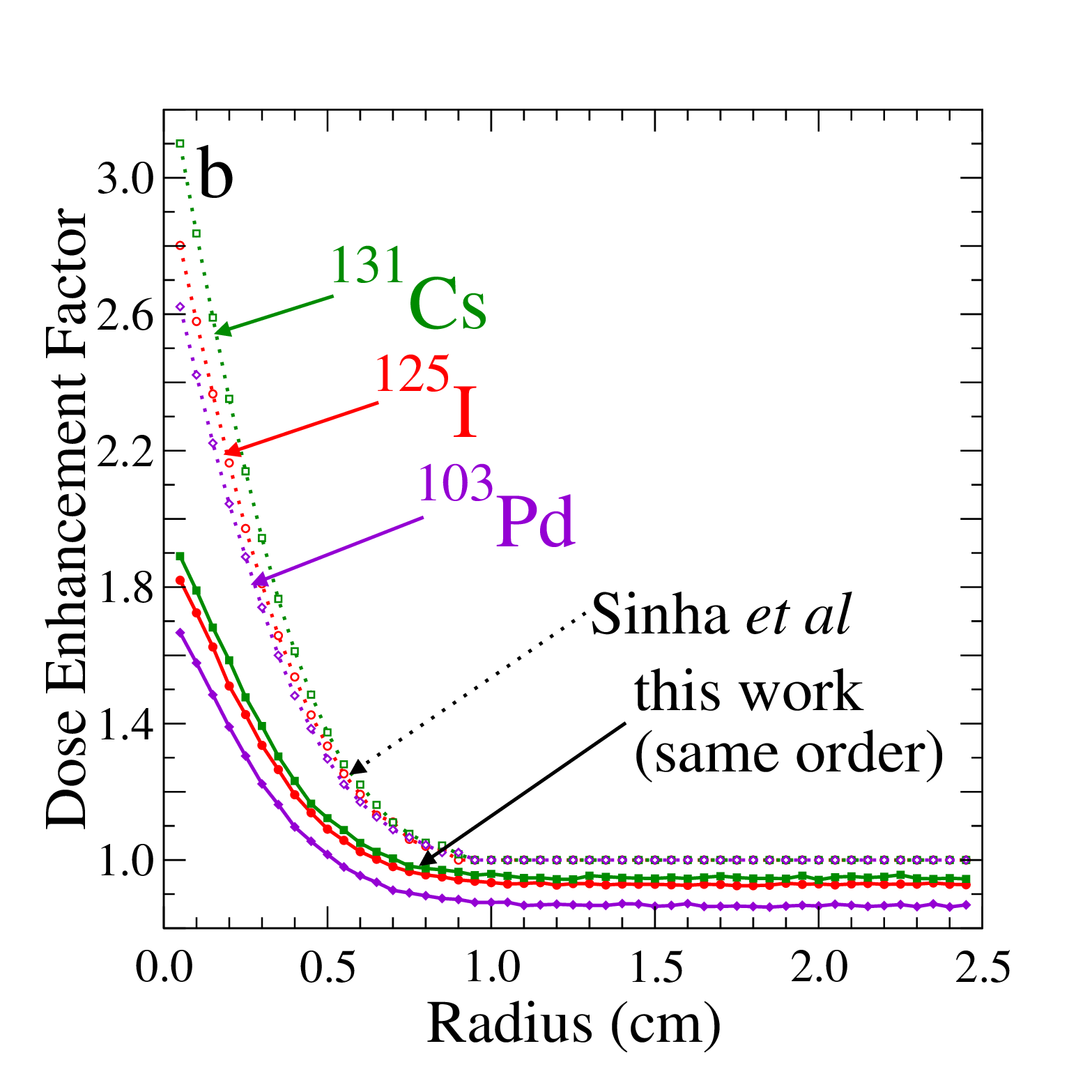}
		\end{subfigure}
	\captionl{(a) DEF versus radius within the spherical phantom (Fig.~\ref{fig:DiagramSpheres}) at different times during treatment with $^{125}$I.  (b) Comparison of DEF versus radius for different radionuclides after 33~days.  Both panels present results of the current work (closed symbols and solid lines; error bars for statistical uncertainties are often too small to see) as well as data extracted from Sinha {\it et al}\cite{Si15} using plot digitization (open symbols and dotted lines). \label{fig:Spheres}}
\end{figure}

DEFs as a function of radius computed within the \acro model for the spherical phantom with varying gold concentration (Fig.~\ref{fig:DiagramSpheres}) are shown in Fig.~\ref{fig:Spheres}a\mfig{fig:Spheres} for $^{125}$I at different times.  Near the source, the DEFs range from 1.4 after 1~day to near 1.9 after 33~days or more.  DEFs go to unity at small radii for the 1 and 5~day cases, whereas decreases in DEFs to unity happen at larger radii later in the treatment (33~days or more).  For each treatment time considered, DEFs drop below one at some radius, \eg 0.35~cm for 5~days, 0.7~cm for 33~days, and 1.2~cm for 200~days.  As observed in the cylindrical phantom computations (section \ref{res:cylinder}), DEFs$<1$ correspond to a dose decrease due to decreasing fluence resulting from gold between the source and scoring regions.  For the current spherical geometry involving varying gold concentrations (Fig.~\ref{fig:DiagramSpheres}b), dose decreases are observed in regions where there is a non-zero concentration of gold, as well as at larger radii where effectively no GNPs have diffused.  Similar trends are observed for the $^{131}$Cs and $^{103}$Pd sources (Fig.~\ref{fig:Spheres}b), however, exact DEF values depend on radionuclide source spectrum, with lower (higher) DEFs for $^{103}$Pd ($^{131}$Cs) compared to $^{125}$I.

DEFs from Sinha \textit{et al}\cite{Si15} are presented alongside the \acro results in Fig.~\ref{fig:Spheres}.  While both sets of DEFs follow the same trends, they differ considerably: DEFs computed by Sinha \textit{et al} are consistently higher for all radii and times, and for all three radionuclides.  For early times in the treatment (1 and 5 days), 40\% discrepancies are observed near the source, but agreement is better near a radius of 5~mm where little gold has diffused.  For later times (33 days or more), differences range from 5 to 50\%.  Most notably, DEFs computed by Sinha \textit{et al} never drop below one -- all DEFs converge to unity with increasing radius whereas the effects of increased fluence perturbation with radius (due to gold), seen in \acro results in Fig.~\ref{fig:Spheres}, are not apparent.

%  _____  _                        _             
% |  __ \(_)                      (_)            
% | |  | |_ ___  ___ _   _ ___ ___ _  ___  _ __  
% | |  | | / __|/ __| | | / __/ __| |/ _ \| '_ \ 
% | |__| | \__ \ (__| |_| \__ \__ \ | (_) | | | |
% |_____/|_|___/\___|\__,_|___/___/_|\___/|_| |_|
\section{Discussion}

The cylindrical and spherical phantom examples illustrate the main ideas of the \acro model for GNPT: distinct models are employed on different length scales in order to capture relevant physics effects within a single, relatively efficient, simulation.  While these ideas are generalizable to consider many issues in GNPT using the \acro approach (more below), the current work focuses on relatively simple \acro models involving homogenized tissue-gold mixtures (or pure tissue) in the bulk of phantoms (possibly with varying gold concentration) with 50 or 119 microscopic regions containing discretely-modeled GNPs in tissue.  Employing the homogenized, tissue-gold mixture (or pure tissue) within the bulk of the phantom enables realistic representation of scatter conditions and determination of fluence.  The discrete modeling of GNPs within smaller ($\sim$\si{\micro\metre}) regions enables scoring of energy deposition within the tissue in which GNPs are embedded.  

The lattices employed herein for modeling arrays of GNPs are not {\it required} as part of the \acro approach, rather, they may be a useful approach for geometry specification.  Lattices have been employed elsewhere for GNPT (discussed above; Refs.~\cite{Me13,Gh12,Zh09,Ca13,Gh13,To12b}), but they have not been combined with the homogenized tissue-gold mixtures as done within our \acro models, nor used within more sophisticated phantoms with multiple GNP lattices to model varying gold concentrations (spherical phantom simulations, section~\ref{res:sphere}).  Furthermore, the \acro approach enables considerably more efficient simulations than otherwise possible with discrete modeling of GNPs throughout the entire GNP-containing volume (as done previously \cite{Me13,Gh12,Zh09,Ca13,Gh13,To12b}).  We observed efficiency gains of factors of up to 122 within the cylindrical phantom (table~\ref{tab:eff}), with efficiency gains varying with source energy and gold concentration.  In general, efficiency gains would depend on the details of the geometry considered, \eg phantom size; number, size, and geometries of scoring regions; GNP concentration, source energy, and so on. In certain circumstances, discrete modelling of each GNP within the macroscopic volume would be impossible with available computer resources (memory, computing time).  Indeed, calculation times for the simulations presented herein would have been prohibitively long without the \acro approach -- performing the current study using discrete modeling of every GNP within the phantoms would have been computationally infeasible.  {\it All} of the cylindrical phantom simulations were completed in less time than it would take to run the cylindrical simulation modeling discrete GNPs everywhere for \textit{only} an energy of 20~keV and concentration of 20~mg/g.  
%\comment{In all my simulations, this case is the slowest by far for HetMS.  Since HetMS is 120 times faster, and we only perform 6*3*2=32 (energies*concentrations*GNPradius) different HetMS sims, I think that last sentence is safe to say.}
 
%\comment{Need to consider whether to put our longer discussion of our results in the context of the literature (for cylinder results) in the previous paragraph (and NOT in the results section).  Could have this come in after the first sentence -- refer to other works to comment on comparable DEFs, and gradients in DEFs.  Could then comment that while other works have similar DEFs and have also observed DEF gradients, none have observed DEFs$<1$ wihtin the treatment volume (region containing DEFs) -- then go back to current sentences where we comment that need to have such effects accounted for.} 

%ROWAN -- good version of comparison paragraph.  
DEFs computed for the cylindrical phantom scenario (section \ref{res:cylinder}) demonstrate the competing effects of enhanced local energy deposition and decreasing fluence with depth in the phantom.  While dose enhancements are observed for many source energies, DEFs change considerably with depth: DEFs$<1$ are observed for lower energy sources (20, 30~keV) within the phantom.  These dose decreases have the potential to create cold spots in a treatment volume and must be accounted for in any algorithm for GNPT dosimetry or treatment planning.

The results of our cylinder simulations are generally comparable with those presented in other published works.  For example, the magnitudes of DEFs observed at the front of the cylindrical phantom for 20 and 30~keV sources (Fig.~\ref{fig:PDD}) are generally in agreement with those predicted by Roeske {\it et al} for $^{103}$Pd and $^{125}$I using analytic calculations \cite{Ro07}.  %[GNP self-absorption of photoelectrons and overestimation of DEFs computed within homogenized models \cite{Le11,Gh12}.  What to say?   Keep references in results section and omit here? Yes, do this for now.]
%\comment{I can't access the library proxy right now, but I thought it was mentionned in one or two places, one of which might have been this paper (http://www.sciencedirect.com/science/article/pii/S1120179715001180) but I'm not sure (can't check).  Could we maybe massage the wording to be a bit less absolute?}, 
Results of some other works show gradients in DEFs over ~$\sim$cm length scales (although often not dose decreases, DEFs$<1$),  \eg MC models employing a $(1$~cm$)^3$ lattice of GNPs and considering kilovoltage photon/brachytherapy sources \cite{Zh09,Gh12}, as well as MC models involving homogenized tissue-gold mixtures (no discrete modelling of GNPs) irradiated by $^{125}$I, 50 kVp x ray, and $^{169}$Yb sources \cite{Ch09d}.  In contrast, results of the similar work of Mesbahi {\it et al} (who modeled a $(1$~cm$)^3$ region containing a lattice of GNPs (of varying diameter; concentrations of 7 and 18~mg/g) within a larger phantom) do not have a downward trend in DEFs with depth.  This is likely due to the relatively low resolution of their dose scoring grid, the shorter distance (1~cm) they considered within the region containing GNPs (along with the magnitude of statistical uncertainties), or the fact that the lattice had an axis aligned with the parallel beam incident on the phantom (see Fig.~1 in Mesbahi {\it et al} \cite{Me13}).  Within the context of tumor dose enhancement using modified megavoltage beams and gadolinium or iodine contrast media (not GNPT), Robar {\it et al} noted decreasing trends in dose enhancement over centimeter length scales \cite{Ro12a}.

With the potential for considerable DEFs at kilovoltage energies, previous studies have investigated DEFs for brachytherapy involving GNPs (see, \eg Refs.~\cite{Ch09d,Jo10,Le11,Gh12,Zh09,Zy13}).  Recent works considered new avenues for GNP delivery within the context of accelerated partial breast irradiation with an electronic brachytherapy source\cite{Ci15} and prostate brachytherapy\cite{Si15} with GNPs diffusing within the treatment volume.  Our spherical phantom simulations considered the latter scenario, with considerable discrepancies (up to 50\%) observed between the published DEFs computed by Sinha {\it et al} and those from our \acro model (Fig.~\ref{fig:Spheres}).  

Sinha \textit{et al} indicate 10\% uncertainties on their DEFs \cite{Si15}, citing another work dealing with dose enhancement to endothelial cells via megavoltage x rays and GNPs \cite{Be11b}, however, it is unclear where the figure of 10\% originates given the different considerations for the kilovoltage sources in the brachytherapy context.  Uncertainties of 10\% do not account for the observed discrepancies in Fig.~\ref{fig:Spheres}.  While seed models and spectra of Sinha {\it et al} were unspecified, different spectra do not explain discrepancies as we found relatively small variations in DEFs computed assuming different spectra of the same radionuclide. % (\eg bare source or average emitted photon energy)
Sinha \textit{et al} further note that their analytic approach\cite{Ng10} to compute DEFs has been validated against MC results in previous publications \cite{Si15}.  However, there are many approximations in this analytic calculation that are questionable in the current context, \eg ignoring the effects of scatter and attenuation of primary photons due to gold.  %using a simple CSDA approximation for photoelectrons generated in gold and <--(RT) I don't understand this without more detail and hence probably better to omit
%neglecting self-absorption of photoelectrons \comment{Looking at this, I vaguely remember Ngwa accounting for the energy lost in the GNP.  I'll check in the morning, I can't access the library proxy right now.  Even if it is, I can come up with other things done wrong}. 
 These approximations may result in the considerable discrepancies between the DEFs computed via the analytic approach and the \acro model.  The validity of the assumptions of this analytic approach\cite{Ng10} across the diverse published applications\cite{Be11b,Si15,Ng12b,Ci15} is doubtful.

Validation of computational models used to determine DEFs is generally challenging due to the lack of experimental data for comparison and difficulties in cross comparisons between distinct published (computational) works with their differences in geometries, modeling assumptions, sources, {\it etc}.  For the current work, with the alterations to egs\_chamber and the creation of the lattice geometry, the code was tested for self-consistency.  Simulations were tested for robustness under variations in transport parameters.  The use of 1~keV energy transport cutoffs for photons and electrons was investigated by repeating simulations with these cutoffs increasing in 125~eV increments up to 2~keV (\ie 1.000~keV, 1.125~keV, 1.250~keV, ..., 2.000~keV); results agreed within statistical uncertainties up to 1.75~keV.  Several results in Fig.~\ref{fig:PDD} were replicated using 10~\si{\micro\metre} long scoring cylinders instead of 100~\si{\micro\metre} cylinders, hence demonstrating insensitivity to variations in the \si{\micro\metre} scoring region dimensions.  

The demonstrated agreement between our results and those of Koger and Kirkby (Table \ref{tab:Koger}, section \ref{res:KK16}) from their PENELOPE simulations (100~eV transport cutoff) provides validation of our simulations, demonstrating the appropriateness of EGSnrc with its condensed history (CH) approach (class II scheme\cite{Ka99a}), as well as our 1 keV transport cutoff and lattice geometry, for the scenarios considered herein.  Cai \textit{et al} noted good agreement when comparing their MCNP5 (CH algorithm for electrons, 1~keV transport cutoffs) results within $\mu$m-scale cell compartments (cell and nucleus diameters of 18.6~$\mu$m and 12.6~$\mu$m, respectively) \cite{Ca13} to those previously published using PENELOPE\cite{Le11} (50 eV electron cutoff) when scoring in microscopic regions.  These observations support the use of CH codes and $\sim1$~keV transport cutoffs in GNPT scenarios in which energy deposition need not be resolved on nm length scales (40 nm range of 1 keV electron in water \cite{Me02c}) and when self-absorption of sub-1~keV electrons within GNPs \cite{Le11c,Le11} is relatively unimportant.  %relative to the overall energy deposition \cite{Le11c,Le11}.  
While use of such CH codes and 1~keV transport cutoffs enhance simulation efficiency, thus making simulations possible with available computing power, some GNPT applications require consideration of electron transport to lower energies and event-by-event simulation of electron transport (with various associated challenges\cite{Ni06,TK11,ZS16}).  These applications include quantification of energy deposition with $\sim$nm resolution in the immediate vicinity of GNPs towards understanding possible considerable DEF variability on nm length scales\cite{Zy13,Li14a}.  

%Some of our results in Fig.~\ref{fig:PDD} should be comparable to those of Mesbahi {\it et al}\cite{Me13} who investigated DEFs in a $(1$~cm$)^3$ region containing a lattice of GNPs (varying diameter) within a larger phantom, however gold concentrations and GNP diameters differ, and the particular photon incident energies differ (with some overlap).  Comparing the results for 20~nm GNPs at 5~mg/g (Fig.~\ref{fig:PDD}a) to those of Mesbahi {\it et al} with 30~nm GNPs at 7~mg/g, both for an incident 50~keV beam, DEFs are near 1.7-1.8 up to 1~cm within the GNP region.  However, there are discrepancies at higher concentrations: still considering 50 keV photons but gold higher concentrations, the DEFs of Mesbahi {\it et al} for 18~mg/g do not show the downward trend with depth evident in Figure~\ref{fig:PDD}b, with DEFs staying roughly constant as a function of depth.  These discrepancies may be due to the relatively low resolution of their dose scoring grid, the shorter distance (1~cm) they considered within the region containing GNPs (along with the magnitude of statistical uncertainties), or the fact that the lattice has an axis aligned with the parallel beam incident on the phantom (see Fig.~1 in Mesbahi {\it et al} \cite{Me13}).

In their recent review article\cite{ZS16}, Zygmanski and Sajo emphasized the importance of multiscale considerations for GNPT.  The relevant discussion regarding simulation mainly focused on a two-stage approach, \eg first calculating a phase space within an efficient macroscopic phantom and then creating a microbeam in a more detailed microscopic phantom (see Ref.~\cite{ZS16} and references therein).  This approach involves two separate simulations and sometimes different macro/micro codes; typically, only a few depths within a phantom are considered\cite{Zy13,Le11,KG15}.  However, as seen in Figs.~\ref{fig:PDD}-\ref{fig:Spheres}, variations in DEFs with depth may be considerable and of importance for prospective clinical GNPT scenarios; indeed, variations in DEFs with depth may be more considerable than much-studied variations of DEFs with source energy or gold concentration (see Ref.~\cite{ZS16} and references therein).  In contrast to the two-stage approach, the \acro concept enables a single relatively-efficient simulation modeling both macro- and microscopic geometries: energy deposition may be computed within multiple microscopic regions throughout the macroscopic phantom while still accounting for scatter conditions.

The \acro approach described herein may be both implemented and applied in diverse ways for GNPT.  
%implementation
While we implemented the \acro model within EGSnrc, the ideas are not limited to one code system and might be implemented in different codes.  Advanced geometry modeling packages/capabilities would be a requirement of such a code; any efficiency gains would depend on the code details including geometry modeling approach, variance reduction techniques, etc.  Code choice would also be influenced by the application considered, \eg whether simulation of radiation transport to sub-1~keV energies is needed.  Such considerations are required in the investigation of interactions of radiation in the vicinity of DNA and GNPs\cite{Ja11} or quantification of highly localized (nm) dose enhancement in the immediate vicinity of a GNP \cite{Jo10,Le11,Zy13}. Radiation transport modeling could be different within distinct parts of the \acro model, \eg lower transport cutoffs and distinct transport algorithms within the microscopic geometry.
%(see earlier discussion - refs).  %might be useful for considering energy deposition on nm length scales, as described above \cite{Zy13} [add ref to Lin et al, PMB vol 59 (2014) pp7675-7689]. \comment{last sentence may be too repetitive}
Diverse GNPT scenarios including different possible radiotherapy sources \cite{Li14a,Ro07} could also be considered with the \acro approach, computing various quantities of interest beyond tissue cavity DEFs and nuclear DEFs, \eg spectra of radiation quanta within different regions or emitted from GNPs\cite{Le11}, enhanced energy deposited within mitochondria\cite{KG15,Mc16}.  

While our example simulations generally involved two distinct scales (1 -- homogeneous gold-tissue mixtures; 2 -- discrete GNPs embedded in tissue), there could be more than two `levels' of detail: different models on multiple scales in order to accurately model radiation transport and energy deposition.   Larger and more detailed phantoms might be considered.  Microscopic regions could have varying and more detailed arrangements of GNPs within cells\cite{CC07,Le11} according to biological distribution (including possible clustering), with  subcellular structures included, \eg nucleus, mitochondria, DNA \cite{KG15,Do13,Mc16}, towards understanding the biological effects of such treatments.  These studies, coupled with work on radiobiological effect, might help in exploring unanswered questions within the field of GNPT, \eg decreased cell survival in GNP treatments in the megavoltage range\cite{Ch10b,Ja11}.

\section{Conclusion}

The current work presents a simultaneous multiscale approach, described as the \model (\acroc) model, accounting for both macroscopic and microscopic considerations in a single simulation to accurately model the effects of GNPs within the radiotherapy context.  We have demonstrated the \acro approach within the simple model of a cylindrical phantom consisting of homogeneous gold-tissue mixture in which microscopic regions of discretely-modeled GNPs in tissue are embedded, as well as within a recently-proposed prostate brachytherapy scenario involving GNPs diffusing from spacers.  The \acro approach enables consideration of both scatter conditions and changes in fluence over the averaged gold-tissue medium, as well as scoring in tissue containing GNPs but not within the GNPs themselves.  Compared to discretely-modeling GNPs throughout the gold-containing volume, efficiency gains of factors of up to 122 were observed with \acro simulations, enabling the calculations within the diverse scenarios considered herein with available computing power.  Both sets of example calculations demonstrate considerable and varying dependence of DEFs with depth in tissue, gold concentration, GNP size, and source energy.  These results, as well as observed discrepancies with DEFs computed using a broadly-applied analytic technique involving considerable approximations, emphasize the importance of \acro modeling for lower source energies, \eg brachytherapy or orthovoltage treatments using GNPs.  The \acro MC simulations may be extended in various ways, offering new avenues for GNPT research.

%A majority of the research needed for this work was in comparing different models in the literature and ensuring the validity of the \acro.  Although tedious, it is a necessary step when using MC codes to simulate GNPT; a necessary step many GNPT studies neglect, or even ignore.  Thorough validation does have its advantages as well; a carefully created and extensively tested model is more robust and could be used in more applications.  The \acro could be extended to include more complex geometries in different regions of interest, \eg microscopic regions of interest with varying and more detailed arrangements of GNPs could be used to investigate the effects of varying GNP configuration within a cell \cite{CC07,Do13,KG15}, offering insight into relations between computed DEFs to experimental results for cell survival.

\section*{Acknowledgements}
The authors acknowledge support from the Natural Sciences and Engineering Research Council of Canada (NSERC), Canada Research Chairs (CRC) program, an Early Researcher Award from the Ministry of Research and Innovation of Ontario, and the Carleton University Research Office, as well as access to computing resources from Compute/Calcul Canada and the Shared Hierarchical Academic Research Computing Network (SHARCNET).  The authors declare that they do not have any conflict of interest.

%%%%%%%%%%%%%%%%%%%%%%%%%%%%%%%%%%%%%%%%%%%%%%%%%%%%%%%%%%%%%%%%%%%%%%%%%%%%%%%%%%%%%%%%%%%%%
% BIBLIOGRAPHY
%%%%%%%%%%%%%%%%%%%%%%%%%%%%%%%%%%%%%%%%%%%%%%%%%%%%%%%%%%%%%%%%%%%%%%%%%%%%%%%%%%%%%%%%%%%%%
% Good for drafts
\setlength{\baselineskip}{0.55cm}
% File path to .bib containing references

% File path to .bst containing biblio style
%\bibliographystyle{./medphy}

\end{document}